\documentclass[doublecol]{epl2}
\usepackage{amsmath, amssymb, amsfonts}
\usepackage{graphicx}
\usepackage{bm}
\usepackage{physics}
\usepackage{xcolor}

\definecolor{myblue}{RGB}{0,70,140}

\usepackage[
    colorlinks=true,
    linkcolor=myblue,
    citecolor=myblue,
    urlcolor=myblue
]{hyperref}

\title{Geometry, elasticity, and activity in the transport of self-propelled filaments in turbulence}
\author {Kunal Kumar\inst{1}, Aliv Sahoo\inst{2}, Rahul Kumar Singh\inst{3} and Samriddhi Sankar Ray\inst{1}}
\shortauthor{Kumar, Sahoo, Singh, and  Ray}
\institute{                    
  \inst{1} International Centre for Theoretical Sciences, Tata Institute of Fundamental Research, Bangalore 560089, India\\
  \inst{2} Department of Physics, Indian Institute of Science, Bengaluru 560012, India \\
  \inst{3} Complex Fluids and Flows Unit, Okinawa Institute of Science and Technology Graduate University, Okinawa 904-0495, Japan
}

\abstract{We investigate the transport of elastic active filaments in
two-dimensional turbulence, focusing on how propulsion geometry and elasticity
determine vortex trapping and transport. Using a bead--spring model with
activity applied at the filament head, we compare propulsion that follows the
instantaneous filament conformation with propulsion imposed along a fixed
external direction.We find that activity does not generically enhance
transport: when propulsion remains coupled to the filament backbone, vortex
trapping remains dominant and motion stays effectively diffusive, whereas
fixed-direction propulsion enables persistent excursions across flow structures
and leads to superdiffusive transport. In both cases, activity shifts filament
conformations toward more extended states, effectively opposing elastic
relaxation without eliminating preferential sampling of coherent vortical
regions. At low Weissenberg number, this conformational change is amplified:
activity cooperates with elasticity to enhance preferential sampling of
vortical regions and strengthen vortex trapping. Transport therefore emerges
from a competition between activity, elasticity, and flow-induced deformation,
with elasticity determining how effectively activity-induced extensions can
persist against turbulent trapping. These results establish propulsion geometry
as the key control parameter for transport, with elasticity and activity acting
cooperatively rather than independently to shape filament dynamics in turbulent
flows.} 

\begin{document}
\maketitle

\section{Introduction}

Transport in turbulent flows depends sensitively on the nature of the advected
object~\cite{BecReview,voth2017anisotropic}. Ideal tracers sample the flow uniformly and reflect
the Lagrangian statistics of the carrier flow, whereas inertial or finite-size
particles preferentially concentrate in specific flow regions~\cite{PicardoPRF2019} due to
dissipative dynamics~\cite{bec2006acceleration,
calzavarini2008quantifying,toschi2009lagrangian}. Extended ~\cite{picardo2020dynamics} and deformable
objects ~\cite{verhille2022deformability, brouzet2014flexible} introduce an additional level of complexity because their internal
degrees of freedom couple directly to local velocity gradients. Recent studies
of elastic chains and fibers have shown that extensibility alone can induce
strong preferential sampling and vortex trapping through a coupling between
flow-induced stretching and elastic relaxation
\cite{picardo2018preferential,marchioli2025flexible}. In these systems,
filaments stretch in straining regions and subsequently coil inside vortices,
leading to transport mechanisms fundamentally different from inertial
clustering~\cite{picardo2018preferential}. Related studies on flexible fibers
and elongated particles further demonstrate how deformation, tumbling, and
rotational dynamics modify transport and alignment in turbulence
\cite{verhille20163d,bec2005clustering,Singh2025}. Together, these results
show that elasticity, much like inertia \cite{singh2020elastoinertial}, provides a pathway for breaking
tracer-like behavior in turbulent flows. Dissipative coupling can further lead to sticky elastic collisions, 
altering encounter statistics and clustering~\cite{sticky}.

Filamentous structures are ubiquitous in biological and synthetic active
systems. Microorganisms such as sperm cells and bacteria propel themselves
using flexible flagella~\cite{lauga2009hydrodynamics}, while cytoskeletal
filaments driven by molecular motors generate active stresses and flows inside
cells~\cite{sanchez2012spontaneous,shelley2016dynamics}. Inspired by such
systems, a wide variety of artificial microswimmers, including catalytic Janus
particles, magnetic filaments, and flexible swimmers driven by external fields,
have been developed for controlled transport in fluid
environments~\cite{ebbens2010pursuit,bechinger2016active}. Unlike passive
particles, these systems generate motion through persistent internal or
externally imposed activity, whose strength and geometry can be tuned.
In parallel, collective active motion in turbulence has been shown to persist or emerge depending on interaction rules 
and driving~\cite{FlockingEPL,FlockingPRF}.

The effect of activity in complex flow environments, however, is not
straightforward~\cite{bechinger2016active}. In many active filament models,
propulsion acts along the filament backbone and therefore remains coupled to
the instantaneous filament configuration
\cite{chelakkot2014flagellar,van2024conformation,duman2018collective}. Since
the configuration itself is continuously shaped by the surrounding flow,
activity does not necessarily enhance transport or dispersion
\cite{qin2022confinement}, particularly in the presence of coherent vortical
structures that promote trapping~\cite{durham2013turbulence,
tarama2014deformable}. Activity therefore competes simultaneously with elastic
relaxation and flow-induced deformation, raising the question of how it alters
the transport mechanisms already present in passive elastic systems.

Previous studies of active filaments have mainly focused on how activity and
elasticity determine filament conformations in quiescent or self-generated
flows~\cite{isele2015self,chelakkot2014flagellar,
sanchez2012spontaneous,dreyfus2005microscopic}. These works show that activity
can drive directed motion, coiling, spirals, buckling, and other nontrivial
configurations depending on the geometry of propulsion
\cite{bianco2018globulelike,prathyusha2022emergent,
tejedor2024progressive}. In particular, the direction of active forcing
relative to the filament backbone strongly influences filament dynamics:
tangential forcing generates compressive stresses and collapse, whereas normal
or polar forcing excites bending and asymmetric deformations
\cite{van2024conformation,bianco2018globulelike,
prathyusha2022emergent,tejedor2024progressive}. These studies establish
propulsion geometry as a key control parameter for filament conformation and,
potentially, transport.

In contrast, the dynamics of active elastic filaments in externally imposed
turbulent flows remain comparatively unexplored. In such environments,
propulsion geometry, elasticity, and flow-induced deformation act
simultaneously, leading to a competition between activity-driven motion and
vortex trapping that is not captured by existing models
\cite{yang2008cooperation,winkler2017active}. It therefore remains unclear
under what conditions activity enhances transport, and whether this depends on
how propulsion is geometrically coupled to the filament configuration.

Here, we address this problem by studying active elastic filaments advected by
two-dimensional turbulence. We compare two propulsion mechanisms: one in which
activity follows the local filament orientation, and another in which
propulsion acts along a fixed external direction. We show that activity alone
does not generically enhance transport. When propulsion remains coupled to the
filament backbone, motion stays strongly correlated with vortex trapping and
dispersion increases only weakly. In contrast, fixed-direction propulsion
partially decouples motion from the instantaneous filament configuration,
allowing excursions from vortical regions and leading to enhanced transport.
In both cases, activity drives filaments toward more extended configurations,
effectively opposing elastic relaxation without eliminating preferential
sampling of flow structures. This effect is amplified for stiffer filaments, where activity-induced extensions persist long enough to enhance preferential sampling of vortical regions. Transport therefore emerges from an interplay
between propulsion geometry, elasticity, and turbulent flow structures, with propulsion geometry controlling transport while elasticity shapes the persistence of filament deformation.

\section{Model and numerical simulations}

\begin{figure*}
    \includegraphics[width=\textwidth]{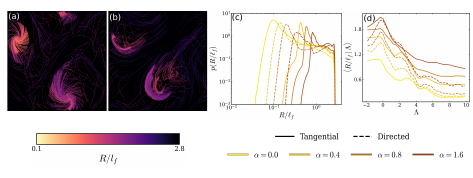}
	\caption{Snapshots of filaments with (a) tangential and (b) directed propulsion, colored by filament length. Tangentially propelled filaments appear more spatially dispersed, whereas directed filaments exhibit partial global alignment along the imposed propulsion direction; we refer the reader to Ref.~\cite{anim_fil} for an animation 
	illustrating this effect. In both cases, coherent collective structures emerge near vortical regions, distinct from the underlying flow streamlines.
(c) Probability density function of the normalized filament length $R/\ell_f$ for different activity strengths $\alpha$. Increasing activity shifts the distribution toward larger filament lengths while preserving its overall shape. Tangentially propelled filaments exhibit systematically larger mean extensions than directed filaments.
(d) Conditional mean filament length $\langle R|\Lambda \rangle$ as a function of the Okubo--Weiss parameter $\Lambda$. Filaments remain more coiled inside vortical regions ($\Lambda>0$), while activity enhances filament extension across all flow regions without eliminating preferential sampling.}
\label{fig:snap}
\end{figure*}

We study an elastic filament advected by a two-dimensional turbulent flow and
driven by an active force applied at its head. The model builds on the elastic
chain formulation introduced in Ref.~\cite{picardo2018preferential}, with
activity incorporated at the level of a single bead.

The background flow is governed by the incompressible, unit-density, two-dimensional
Navier--Stokes equations,
\begin{equation}
\partial_t \mathbf{u} + \mathbf{u}\cdot\nabla\mathbf{u}
=
-\nabla p + \nu \nabla^2 \mathbf{u} + \mathbf{f},
\qquad
\nabla\cdot\mathbf{u}=0,
\end{equation}
where $\mathbf{u}(\mathbf{x},t)$ is the velocity field. The flow is driven to
a statistically steady state by a large-scale forcing
$\mathbf{f}=-F_0k_f\cos(k_fx)$, which injects energy at the scale
$\ell_f=2\pi/k_f$. 

For two-dimensional turbulence, the flow geometry is characterised 
most conveniently via the Okubo--Weiss parameter 
$\Lambda= (\omega^2-\sigma^2)/4 \langle \omega^2 \rangle$, where
$\omega$ is the vorticity and $\sigma$ is the strain rate. This definition leads to $\Lambda > 0$ identifying a vortical region and 
$\Lambda < 0$ a straining region and allows us to measure the flow properties 
in an Eulerian or  Lagrangian framework along particle trajectories,

We solve the Navier--Stokes equations in vorticity form using a pseudospectral
method on a doubly periodic domain of size $L=2\pi$ with resolution
$512^2$. Spatial derivatives are computed in Fourier space using standard
$2/3$ dealiasing, and time integration is performed with a second-order
Runge--Kutta scheme with time step $\Delta t=5\times10^{-4}$. The flow is
forced at large scales with forcing scale $\ell_f\simeq1.25$, corresponding to
the typical vortex size. We use viscosity $\nu=10^{-4}$ and large-scale
friction $\mu=2\times10^{-3}$ to obtain a statistically steady state with a mean kinetic 
energy $E$.

The filament consists of $N_b$ inertia-less beads with positions
$\{\mathbf{x}_i\}$ ($i =  0, \ldots, N_{b} -1$)  connected by nearest-neighbour elastic links. Its configuration is
described by the center of mass
$\mathbf{X}_c=(1/N_b)\sum_i \mathbf{x}_i$ and bond vectors
$\mathbf{r}_j=\mathbf{x}_{j+1}-\mathbf{x}_j$. In the absence of activity, the
filament is advected and stretched by the flow while relaxing through elastic
forces. The bond dynamics is governed by finitely extensible nonlinear elastic
(FENE) springs,
\begin{equation}
f_j=\left(1-\frac{|\mathbf{r}_j|^2}{r_m^2}\right)^{-1},
\end{equation}
which prevent unbounded stretching. 

Activity is introduced through a propulsion velocity of magnitude $v_0$ acting on
the head bead. In what follows, we work with an effective, non-dimensional activity parameter 
$\alpha = v_0/\sqrt{2E}$.
In the first mechanism, the propulsion  direction
follows the local filament orientation of the head segment,
\begin{equation}
\mathbf{p}_0=-\frac{\mathbf{r}_0}{|\mathbf{r}_0|},
\qquad
\mathbf{r}_0=\mathbf{x}_1-\mathbf{x}_0,
\end{equation}
so that activity remains coupled to the instantaneous filament configuration.
In the second mechanism, propulsion acts along a fixed external direction,
$\mathbf{p}_0=\hat{\mathbf e}$ (here $(\hat{\mathbf x}+\hat{\mathbf y})/\sqrt{2}$), independent of filament shape. These two
cases represent fundamentally different couplings between activity and
filament conformation.

The filament center-of-mass dynamics is then
\begin{equation}
\dot{\mathbf X}_c=
\frac{1}{N_b} \biggl(\sum_{i=0}^{N_b-1}\mathbf u(\mathbf x_i,t)
+{v_0}\mathbf p_0
	+\sqrt{\frac{r_{\rm eq} ^2}{2\tau}}
\sum_{i=0}^{N_b-1}\boldsymbol\xi_i(t) \biggr),
\end{equation}
where $\boldsymbol\xi_i(t)$ are independent Gaussian white noises. 
The parameter $\tau$ sets the elastic
relaxation time, with the corresponding ~\cite{jin2007dynamics} chain relaxation time
$\tau_{\rm chain}=[(N_b+1)N_b/6]\tau$. 
This allows us to define the Weissenberg numbers Wi = $\tau_{\rm chain}/t_f$ of our filaments, where the 
flow time-scale $t_f = \ell_f/\sqrt{2E}$ is the typical turnover time of a vortex. 
The equilibrium length of an 
individual bond $r_{\rm eq}$ is set by the noise term.
The bond
dynamics contains the corresponding elastic and active contributions, with the
activity entering only at the head bead.

Filaments are evolved simultaneously with the flow after it has reached a steady state. We simulate
$5\times10^4$ filaments, each with $N_b=10$ beads, maximum bond length
$r_m\simeq0.4$ and equilibrium length scale $r_{\rm eq}=0.01$. We choose relaxation times
$\tau=0.15$ and $\tau = 0.0375$ to achieve Weissenberg numbers Wi = 2.8 and 0.7, respectively, and 
$0 \leq \alpha \leq 3.5$.
Fluid velocities at bead positions are obtained using bilinear
interpolation, and periodic boundary conditions are imposed on both the flow
and filament positions.

\section{Results}

We first examine how activity modifies filament conformations when propulsion
is coupled to the local filament orientation. In this tangential propulsion
case, activity remains slaved to the instantaneous filament configuration, so
that the surrounding flow continuously shapes both filament structure and
propulsion direction. This coupling leads to a competition between
flow-induced trapping and activity-driven extension.

The effect of this competition is illustrated in
Fig.~\ref{fig:snap}(a) for Wi = 2.8 and $\alpha  = 3.2$. Passive filaments remain compact and
closely aligned with coherent flow structures, reflecting
the strong tendency of turbulent vortices to trap and coil
elastic chains~\cite{picardo2018preferential}. Once activity is introduced, filaments
become noticeably more extended. The
head bead persistently attempts to move outward along
the local propulsion direction, while trailing segments
remain pinned by the vortical flow. As a result,
activity does not destroy the organization imposed by the
flow, but instead stretches filaments within these trapped
states.

\begin{figure*}
\includegraphics[width=\textwidth]{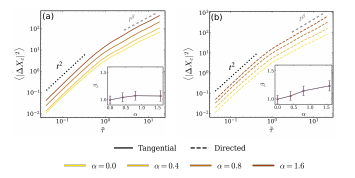}
\caption{Mean-squared displacement of the filament center of mass, versus $\tilde{\tau}$, 
	for (a) tangential and (b) directed propulsion (shifted for visual clarity). Insets show the corresponding transport exponent $\beta$. Tangential propulsion produces only weak enhancement of transport, with the measured exponent remaining close to the diffusive value within error bars. In contrast, directed propulsion leads to substantially stronger superdiffusive transport by enabling larger excursions across flow structures.}
\label{fig:MSD}
\end{figure*}

This change in conformation is quantified by the
probability distribution of the normalized filament length
$R/\ell_f$, where $R = \sum r_j$, shown in Fig.~\ref{fig:snap}(c) (unbroken lines). Increasing
activity shifts the distribution toward larger extensions,
indicating that propulsion systematically opposes elastic
relaxation and maintains filaments in more elongated
configurations. However, the overall shape of the
distribution changes only weakly, suggesting that the
range of accessible filament configurations continues to be
set primarily by the turbulent flow itself. Activity does
not introduce qualitatively new configurations, but rather
biases the dynamics toward more extended realizations of
the same flow-controlled structures.

The persistence of vortex trapping is further evident from
the conditional mean extension
$\langle R|\Lambda\rangle$, shown in
Fig.~\ref{fig:snap}(d) with the unbroken lines, where $\Lambda$ is the
Okubo--Weiss parameter. Filaments remain systematically
more coiled inside vortical regions ($\Lambda>0$) than in
straining regions, even at finite activity. At the same
time, increasing activity enhances extension across all
values of $\Lambda$. Activity therefore modifies how
filaments deform within the flow without eliminating the
preferential sampling mechanisms already present in the
passive elastic system.

The filament conformation in the tangential propulsion case is contrasted in Fig.~\ref{fig:snap}(b) with that obtained under externally imposed propulsion at the same Wi. Qualitatively, a comparison of panels (a) and (b) suggests weaker vortex pinning and stronger global ordering along $\hat{\mathbf e}$. This reflects in the  normalized filament length
distribution, Fig.~\ref{fig:snap}(c) (broken lines) which is significantly
shifted relative to the tangentially propelled case for the same values of
$\alpha$ and hence the mean length of filaments are larger with tangential
propulsion.  The contrast between the persistence of vortex trapping, as shown
in Fig.~\ref{fig:snap}(d), is subtle. In the straining region ($\Lambda < 0$)
where filaments stretch, the distinction between tangential and directed
propulsion becomes weaker with increasing $\alpha$. Contrastingly, in vortical
regions ($\Lambda > 0$) the competition between coiling and propulsion
naturally leads to an asymmetry between the two activity mechanisms for all 
$\alpha$. This heuristic picture is borne out quite clearly in
Fig.~\ref{fig:snap}(d) when comparing the broken and unbroken lines.  We refer
the reader to an animation of the evolution of filaments for $\alpha = 0.8$,
in Ref.~\cite{anim_fil}, which clearly contrasts the dynamics of tangential and
directed filaments.

These conformational changes have important consequences for transport.
Activity primarily modifies filament conformations, whereas transport
enhancement depends on whether these conformational changes can overcome vortex
trapping. To characterize filament motion through the turbulent flow, we
examine the mean-squared displacement $\langle |\Delta X_c|^2 \rangle$ of the
filament center of mass. We show loglog plots, for Wi = 2.8, of $\langle
|\Delta X_c|^2 \rangle$ in Fig.~\ref{fig:MSD} for (a) tangential and (b)
directed propulsion as a function of the nondimensional time
$\tilde{\tau}=t/\tau$.  At short times, both propulsion mechanisms exhibit
ballistic behavior, reflecting persistent active motion before the influence of
turbulent trapping becomes significant.  Insets show the transport exponent
$\beta$, obtained from a local slope analysis of the MSD curves in the
non-ballistic regime; symbols denote the mean value of $\beta$, while error
bars indicate the corresponding standard deviation. For tangential propulsion,
activity enhances the overall dispersion, but the measured exponent remains
consistent with diffusive transport $\beta = 1.0$ within error bars. This
behavior reflects the fact that activity primarily stretches and reorients
filaments within trapped states rather than enabling sustained escape from
vortical regions. Since propulsion continuously follows the instantaneous
filament shape, the active forcing remains dynamically tied to the same flow
structures that generate trapping. In contrast, directed propulsion produces a
clear superdiffusive regime at large $\alpha$, consistent with more persistent
excursions across coherent flow structures.

\begin{figure}
    \centering
    \includegraphics[width=0.95\linewidth]{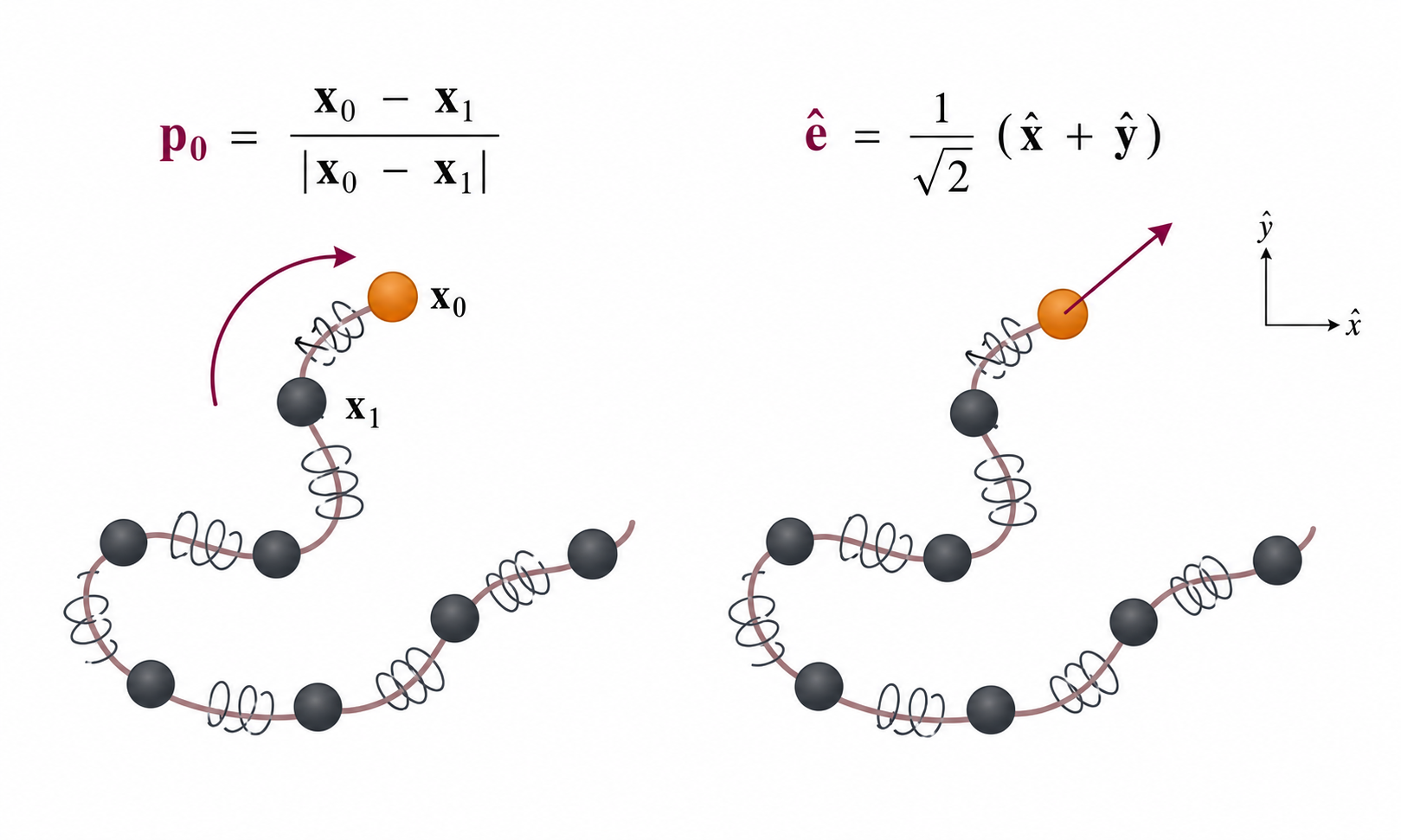}
    \caption{Schematic of the two propulsion mechanisms considered for active filaments. The filament consists of beads connected by elastic links, with activity applied at the head bead (orange). Left: tangential propulsion, where $\mathbf{p}_0 = -(\mathbf{x}_1-\mathbf{x}_0)/|\mathbf{x}_1-\mathbf{x}_0|$ follows the local filament orientation and remains coupled to the instantaneous filament conformation. Right: directed propulsion, where $\mathbf{p}_0=\hat{\mathbf e}$ (here $(\hat{\mathbf x}+\hat{\mathbf y})/\sqrt{2}$) is fixed and independent of filament conformation. These two cases illustrate the distinction between geometry-dependent and externally imposed propulsion.
}
\label{fig:schematic}
\end{figure}

This difference can be understood in terms of how
propulsion couples to filament conformation, as
illustrated schematically in Fig.~\ref{fig:schematic}. A
qualitatively different behavior emerges when the
propulsion direction is externally imposed and therefore
partially decoupled from the filament configuration. In
this directed propulsion case, activity no longer follows
the local filament orientation and can act persistently
against the trapping dynamics imposed by the flow.

This weakening of trapping produces a much stronger
effect on transport. Figure~\ref{fig:MSD}(b) shows
substantially enhanced superdiffusive behavior compared
to the tangential case, indicating more effective
exploration of the flow. Because the propulsion
direction is no longer enslaved to the instantaneous
filament shape, active forcing can drive the head bead
persistently across flow structures, allowing filaments
to escape vortices more efficiently and undergo
larger-scale excursions through the turbulent field.

The contrasting transport behavior in
Fig.~\ref{fig:MSD} demonstrates that activity alone is
not sufficient to determine transport properties in
turbulent flows. Instead, the effectiveness of activity
depends crucially on how propulsion couples to filament
conformation. When activity remains geometrically tied to
the filament backbone, vortex trapping continues to
dominate the dynamics. When this coupling is relaxed,
activity becomes more effective at opposing coherent flow
structures and enhancing transport.

\begin{figure}
    \centering
    \includegraphics[width=1.0\columnwidth]{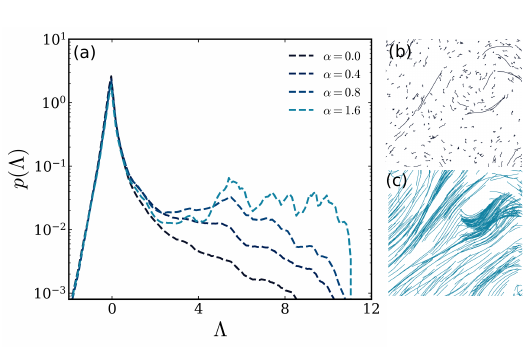}
\caption{(a) Probability density function of the Okubo--Weiss parameter $\Lambda$ sampled along trajectories of 
directed active filaments for $Wi=0.7$. Increasing activity progressively enhances the 
positive-$\Lambda$ tail, indicating stronger preferential sampling of vortical regions.
	Panels (b) and (c) show representative filament configurations for different activity strengths $\alpha$, 
	with colors matched to the corresponding distributions in panel (a). 
	Increasing activity leads to more extended filament conformations and stronger accumulation within vortical regions, illustrating the connection between activity-induced stretching and enhanced preferential sampling.}

	\label{fig:PDF}
\end{figure}

Whether activity ultimately enhances transport or reinforces trapping depends
on the balance between propulsion persistence across flow structures and the
activity-induced extension of the filament, which becomes increasingly
important at lower Weissenberg number.  The results reported so far were for a
moderately large Weissenberg number where, even in the passive $\alpha = 0$
limit, the filaments are already stretched. Activity further enhances this
effect, in different ways, as illustrated in Fig.~\ref{fig:snap}(c). This
suggests that perhaps for small enough Wi, activity can enhance the effective
elasticity of such filaments and make them sample the vortical region more
strongly. A similar tendency for more elastic filaments leading to greater preferential
sampling of $\Lambda > 0$ regions was already reported for  passive ($\alpha = 0$)
filaments in Ref.~\cite{picardo2018preferential}.

This motivates us to examine how elasticity mediates the
competition between activity and turbulent trapping for 
Wi = 0.7.
Figure~\ref{fig:PDF}(a) shows the probability
distribution of the Okubo--Weiss parameter sampled
along directed filament trajectories for different activity
strengths. Increasing activity
progressively enhances the positive-$\Lambda$ tail,
indicating stronger preferential sampling of vortical
regions for the reasons that we discussed above. The corresponding filament configurations shown
in Fig.~\ref{fig:PDF}(b,c), color matched to the
distributions in panel (a), demonstrate that increasing
activity drives filaments toward more extended states
that remain strongly localized within coherent vortical
structures in a manner reminiscent of the phenomenon reported 
by Picardo \textit{et al.}~\cite{picardo2018preferential}. This behavior suggests that activity and
elasticity act cooperatively: activity sustains extended
filament conformations, while elasticity determines how
effectively these deformations can persist within the
flow. As a result, more elastic filaments are able to
maintain activity-induced stretching for longer times,
thereby modifying both trapping and transport
properties.

Taken together, these results show that transport in
active filament systems emerges from a three-way
competition between propulsion geometry, elasticity, and
coherent turbulent structures. Activity primarily acts by
modifying filament conformation and sustaining extended
states, while the geometry of propulsion determines
whether these conformational changes can translate into
effective transport through the flow.

\section{Discussion and conclusion}

We have investigated the
transport of active elastic filaments in two-dimensional
turbulence, focusing on how propulsion geometry and
elasticity shape the competition between activity and
vortex trapping. Our results show that activity alone
does not generically enhance transport. Instead, its
effect depends crucially on how propulsion couples to
the instantaneous filament configuration.

When propulsion follows the filament backbone,
activity remains dynamically slaved to the same flow
structures that trap and coil passive elastic filaments.
In this case, activity primarily modifies filament
conformations, sustaining more extended 
states while leaving the underlying preferential sampling
of vortical regions largely intact. As a result, transport
is enhanced only weakly, since activity mainly drives
stretching and reorientation within trapped states rather
than sustained escape from coherent vortices.

A qualitatively different picture emerges when
propulsion is imposed independently of the filament
configuration. While such filaments also preferentially sample 
the vortices, directed propulsion partially decouples
active forcing from instantaneous filament shape and enables larger-scale
excursions through the flow. The resulting transport
enhancement is therefore geometric in origin: it is not
activity strength alone, but the way activity couples to
filament conformation, that determines whether active
motion can effectively compete with turbulent trapping.

Across both propulsion mechanisms, activity drives
filaments toward more extended configurations. In this sense,
activity acts not only as a source of propulsion, but also
as a mechanism that dynamically maintains extended
filament conformations against turbulent compression.
Elasticity then determines how effectively these extended
states can persist within the flow and therefore how
strongly activity can influence transport. More elastic
filaments sustain larger activity-induced extensions and
consequently exhibit stronger transport enhancement.
Activity and elasticity therefore act cooperatively rather
than independently in shaping filament transport.

A particularly striking consequence of this coupling emerges at low Weissenberg
number, where activity does not merely modify how filaments are transported but
alters their effective elastic character. By sustaining extended conformations
against  elastic relaxation, activity mimics the effect of reduced chain
stiffness, leading to stronger vortex trapping and enhanced preferential
sampling. This cooperative interplay between activity and elasticity --- where
propulsion stabilizes conformations that the flow would otherwise collapse ---
suggests that activity strength and elasticity are not independent control
parameters in such systems, but jointly determine the effective stiffness
experienced by the filament in the turbulent environment.

More broadly, our results place active elastic filaments
within a wider class of systems in which additional
internal degrees of freedom produce departures from
tracer-like transport in turbulence. While inertia,
extensibility, and activity all generate preferential
sampling and anomalous transport, the mechanisms are
fundamentally distinct: activity introduces persistent
self-driven forcing whose effectiveness depends strongly
on geometry and deformation.

These findings open several directions for future work.
In particular, it would be interesting to study
time-dependent or stochastic propulsion strategies, the
behavior of active filaments in three-dimensional
turbulence, and situations in which activity, elasticity,
and inertia coexist simultaneously. More generally, our
results suggest that dynamically tunable propulsion or
effective stiffness could provide a route toward
controlling transport and navigation of active filaments
in complex flow environments.

\begin{acknowledgements}
AS acknowledges support from the Kishore Vaigyanik Protsahan Yojana (KVPY), Department of Science and Technology (DST), Government of India.
SSR acknowledges the Indo–French Centre for the
Promotion of Advanced Scientific Research (IFCPAR/CEFIPRA, project no. 6704-1)
for support.  This research was supported in part by the International Centre
for Theoretical Sciences (ICTS) for the program --- 11th Indian Statistical
Physics Community Meeting (code: ICTS/11thISPCM2026/04). The simulations were
performed on the ICTS clusters Mario, Tetris, and Contra. KK and SSR 
acknowledge the support of the
Department of Atomic Energy, Government of India, under project no.RTI4019 and RTI4013.
\end{acknowledgements}


\begin{thebibliography}{10}

\bibitem{BecReview}
J.~Bec, K.~Gustavsson, and B.~Mehlig.
\newblock Statistical models for the dynamics of heavy particles in turbulence.
\newblock \href{https://doi.org/10.1146/annurev-fluid-032822-014140}
{{\em Annu. Rev. Fluid Mech.}, \textbf{56}, 189 (2024).}



\bibitem{voth2017anisotropic}
G.~A.~Voth and A.~Soldati.
\newblock Anisotropic particles in turbulence.
\newblock \href{https://doi.org/10.1146/annurev-fluid-010816-060135}
{{\em Annu. Rev. Fluid Mech.}, \textbf{49}, 249 (2017).}

\bibitem{PicardoPRF2019}
J.~R.~Picardo, L.~Agasthya, R.~Govindarajan, and S.~S.~Ray
\newblock Flow structures govern particle collisions in turbulence.
\newblock \href{https://doi.org/10.1103/PhysRevFluids.4.032601}
{{\em Phys. Rev. Fluids}, \textbf{4}, 032601 (2019).}

\bibitem{bec2006acceleration}
J.~Bec, L.~Biferale, G.~Boffetta, A.~Celani, M.~Cencini, A.~Lanotte,
  S.~Musacchio, and F.~Toschi.
\newblock Acceleration statistics of heavy particles in turbulence.
\newblock \href{https://doi.org/10.1017/S002211200500844X}
{{\em J. Fluid Mech.}, \textbf{550}, 349 (2006).}

\bibitem{calzavarini2008quantifying}
E.~Calzavarini, M.~Cencini, D.~Lohse, and F.~Toschi.
\newblock Quantifying turbulence-induced segregation of inertial particles.
\newblock \href{https://doi.org/10.1103/PhysRevLett.101.084504}
{{\em Phys. Rev. Lett.}, \textbf{101}, 084504 (2008).}

\bibitem{toschi2009lagrangian}
F.~Toschi and E.~Bodenschatz.
\newblock Lagrangian properties of particles in turbulence.
\newblock \href{https://doi.org/10.1146/annurev.fluid.010908.165210}
{{\em Annu. Rev. Fluid Mech.}, \textbf{41}, 375 (2009).}

\bibitem{picardo2020dynamics}
J.~R.~Picardo, R.~Singh, S.~S.~Ray, and D.~Vincenzi.
\newblock Dynamics of a long chain in turbulent flows: impact of vortices.
\newblock \href{https://doi.org/10.1098/rsta.2019.0405}
{{\em Phil. Trans. R. Soc. A}, \textbf{378}, 20190405 (2020).}

\bibitem{verhille2022deformability}
G.~Verhille.
\newblock Deformability of discs in turbulence.
\newblock \href{https://doi.org/10.1017/jfm.2021.1074}
{{\em J. Fluid Mech.}, \textbf{933}, A3 (2022).}

\bibitem{brouzet2014flexible}
C.~Brouzet, G.~Verhille, and P.~Le~Gal.
\newblock Flexible fiber in a turbulent flow: A macroscopic polymer.
\newblock \href{https://doi.org/10.1103/PhysRevLett.112.074501}
{{\em Phys. Rev. Lett.}, \textbf{112}, 074501 (2014).}

\bibitem{picardo2018preferential}
J.~R.~Picardo, D.~Vincenzi, N.~Pal, and S.~S.~Ray.
\newblock Preferential sampling of elastic chains in turbulent flows.
\newblock \href{https://doi.org/10.1103/PhysRevLett.121.244501}
{{\em Phys. Rev. Lett.}, \textbf{121}, 244501 (2018).}

\bibitem{marchioli2025flexible}
C.~Marchioli, M.~E.~Rosti, and G.~Verhille.
\newblock Flexible fibers in turbulence.
\newblock \href{https://doi.org/10.1146/annurev-fluid-112723-050451}
{{\em Annu. Rev. Fluid Mech.}, \textbf{58}, 167 (2025).}

\bibitem{verhille20163d}
G.~Verhille and A.~Bartoli.
\newblock 3d conformation of a flexible fiber in a turbulent flow.
\newblock \href{https://doi.org/10.1007/s00348-016-2201-1}
{{\em Exp. Fluids}, \textbf{57}, 117 (2016).}

\bibitem{bec2005clustering}
J.~Bec, A.~Celani, M.~Cencini, and S.~Musacchio.
\newblock Clustering and collisions of heavy particles in random smooth flows.
\newblock \href{https://doi.org/10.1063/1.1940367}
{{\em Phys. Fluids}, \textbf{17}, 073301 (2005).}

\bibitem{Singh2025}
R.~K.~Singh.
\newblock Elasticity of fibers prefers the chaos of turbulence.
\newblock \href{https://doi.org/10.1103/PhysRevE.111.L053101}
{{\em Phys. Rev. E}, \textbf{111}, L053101 (2025).}

\bibitem{singh2020elastoinertial}
R.~K.~Singh, M.~Gupta, J.~R.~Picardo, D.~Vincenzi, and S.~S.~Ray.
\newblock Elastoinertial chains in a two-dimensional turbulent flow.
\newblock \href{https://doi.org/10.1103/PhysRevE.101.053105}
{{\em Phys. Rev. E}, \textbf{101}, 053105 (2020).}

\bibitem{sticky}
J. Bec, S. Musacchio, and S.~S.~Ray.
\newblock Sticky elastic collisions.
\newblock \href{https://doi.org/10.1103/PhysRevE.87.063013}
{{\em Phys. Rev. E}, \textbf{87}, 064501 (2013).}

\bibitem{lauga2009hydrodynamics}
E.~Lauga and T.~R.~Powers.
\newblock The hydrodynamics of swimming microorganisms.
\newblock \href{https://doi.org/10.1088/0034-4885/72/9/096601}
{{\em Rep. Prog. Phys.}, \textbf{72}, 096601 (2009).}

\bibitem{sanchez2012spontaneous}
T.~Sanchez, D.~T.~N.~Chen, S.~J.~DeCamp, M.~Heymann, and Z.~Dogic.
\newblock Spontaneous motion in hierarchically assembled active matter.
\newblock \href{https://doi.org/10.1038/nature11591}
{{\em Nature}, \textbf{491}, 431 (2012).}

\bibitem{shelley2016dynamics}
M.~J.~Shelley.
\newblock The dynamics of microtubule/motor-protein assemblies in biology and
  physics.
\newblock \href{https://doi.org/10.1146/annurev-fluid-010814-013639}
{{\em Annu. Rev. Fluid Mech.}, \textbf{48}, 487 (2016).}

\bibitem{ebbens2010pursuit}
S.~J.~Ebbens and J.~R.~Howse.
\newblock In pursuit of propulsion at the nanoscale.
\newblock \href{https://doi.org/10.1039/B918598D}
{{\em Soft Matter}, \textbf{6}, 726 (2010).}

\bibitem{bechinger2016active}
C.~Bechinger, R.~Di~Leonardo, H.~L{\"o}wen, C.~Reichhardt,
  G.~Volpe, and G.~Volpe.
\newblock Active particles in complex and crowded environments.
\newblock \href{https://doi.org/10.1103/RevModPhys.88.045006}
{{\em Rev. Mod. Phys.}, \textbf{88}, 045006 (2016).}

\bibitem{FlockingEPL}
A. Choudhary, D. Venkataraman, and S. S. Ray	
\newblock Effect of inertia on Model Flocks in a Turbulent Environment.
\newblock \href{https://doi.org/10.1209/0295-5075/112/24005}
{{\em Europhys. Lett.}, \textbf{112}, 24005 (2015).}

\bibitem{FlockingPRF}	
A.~Gupta, A. Roy, A. Saha, and S. S. Ray
\newblock Flocking of active particles in a turbulent flow.
\newblock \href{https://doi.org/10.1103/PhysRevFluids.5.052601}
{{\em Phys. Rev. Fluids (Rapids)}, \textbf{5}, 053105 (2020).}

\bibitem{chelakkot2014flagellar}
R.~Chelakkot, A.~Gopinath, L.~Mahadevan, and M.~F.~Hagan.
\newblock Flagellar dynamics of a connected chain of active, polar, brownian
  particles.
\newblock \href{https://doi.org/10.1098/rsif.2013.0884}
{{\em J. R. Soc. Interface}, \textbf{11}, 20130884 (2014).}

\bibitem{van2024conformation}
L.~van~Steijn, M.~Fazelzadeh, and S.~Jabbari-Farouji.
\newblock Conformation and dynamics of wet externally actuated filaments with
  tangential active forces.
\newblock \href{https://doi.org/10.1103/PhysRevE.110.064504}
{{\em Phys. Rev. E}, \textbf{110}, 064504 (2024).}

\bibitem{duman2018collective}
{\"O}.~Duman, R.~E.~Isele-Holder, J.~Elgeti, and G.~Gompper.
\newblock Collective dynamics of self-propelled semiflexible filaments.
\newblock \href{https://doi.org/10.1039/C8SM00438J}
{{\em Soft Matter}, \textbf{14}, 4483 (2018).}

\bibitem{qin2022confinement}
B.~Qin and P.~E.~Arratia.
\newblock Confinement, chaotic transport, and trapping of active swimmers in
  time-periodic flows.
\newblock \href{https://doi.org/10.1126/sciadv.add6196}
{{\em Sci. Adv.}, \textbf{8}, eadd6196 (2022).}

\bibitem{durham2013turbulence}
W.~M.~Durham, E.~Climent, M.~Barry, F.~De~Lillo,
  G.~Boffetta, M.~Cencini, and R.~Stocker.
\newblock Turbulence drives microscale patches of motile phytoplankton.
\newblock \href{https://doi.org/10.1038/ncomms3148}
{{\em Nat. Commun.}, \textbf{4}, 2148 (2013).}

\bibitem{tarama2014deformable}
M.~Tarama, A.~M.~Menzel, and H.~L{\"o}wen.
\newblock Deformable microswimmer in a swirl: Capturing and scattering
  dynamics.
\newblock \href{https://doi.org/10.1103/PhysRevE.90.032907}
{{\em Phys. Rev. E}, \textbf{90}, 032907 (2014).}

\bibitem{isele2015self}
R.~E.~Isele-Holder, J.~Elgeti, and G.~Gompper.
\newblock Self-propelled worm-like filaments: spontaneous spiral formation,
  structure, and dynamics.
\newblock \href{https://doi.org/10.1039/C5SM01683E}
{{\em Soft Matter}, \textbf{11}, 7181 (2015).}

\bibitem{dreyfus2005microscopic}
R.~Dreyfus, J.~Baudry, M.~L.~Roper, M.~Fermigier,
  H.~A.~Stone, and J.~Bibette.
\newblock Microscopic artificial swimmers.
\newblock \href{https://doi.org/10.1038/nature04090}
{{\em Nature}, \textbf{437}, 862 (2005).}

\bibitem{bianco2018globulelike}
V.~Bianco, E.~Locatelli, and P.~Malgaretti.
\newblock Globulelike conformation and enhanced diffusion of active polymers.
\newblock \href{https://doi.org/10.1103/PhysRevLett.121.217802}
{{\em Phys. Rev. Lett.}, \textbf{121}, 217802 (2018).}

\bibitem{prathyusha2022emergent}
K.~R.~Prathyusha, F.~Ziebert, and R.~Golestanian.
\newblock Emergent conformational properties of end-tailored transversely
  propelling polymers.
\newblock \href{https://doi.org/10.1039/D2SM00237J}
{{\em Soft Matter}, \textbf{18}, 2928 (2022).}

\bibitem{tejedor2024progressive}
A.~R.~Tejedor, J.~Ram{\'\i}rez, and M.~Ripoll.
\newblock Progressive polymer deformation induced by polar activity and the
  influence of inertia.
\newblock \href{https://doi.org/10.1103/PhysRevResearch.6.L032002}
{{\em Phys. Rev. Research}, \textbf{6}, L032002 (2024).}

\bibitem{yang2008cooperation}
Y.~Yang, J.~Elgeti, and G.~Gompper.
\newblock Cooperation of sperm in two dimensions: synchronization, attraction
  and aggregation through hydrodynamic interactions.
\newblock {\em arXiv:0810.4680} (2008).

\bibitem{winkler2017active}
R.~G.~Winkler, J.~Elgeti, and G.~Gompper.
\newblock Active polymers---emergent conformational and dynamical properties:
  A brief review.
\newblock \href{https://doi.org/10.7566/JPSJ.86.101014}
{{\em J. Phys. Soc. Jpn.}, \textbf{86}, 101014 (2017).}

\bibitem{jin2007dynamics}
S.~Jin and L.~R.~Collins.
\newblock Dynamics of dissolved polymer chains in isotropic turbulence.
\newblock \href{https://doi.org/10.1088/1367-2630/9/10/360}
{{\em New J. Phys.}, \textbf{9}, 360 (2007).}

\bibitem{anim_fil}
Animation showing the evolution of tangential and directed active filaments in
  2D Navier Stokes flow \url{https://youtu.be/_zzwcsmMjW0}.

\end{thebibliography}



\end{document}